\documentclass[useAMS,usenatbib,usegraphicx]{mn2e}
\usepackage{rotating}

\newcommand{\chan}{{\it Chandra}}
\newcommand{\xmm}{{\it XMM-Newton}}

\title[The SMC Wing Survey]{The {\it Chandra} Small Magellanic Cloud Wing 
Survey - the search for X-ray Binaries}
\author[K. E. McGowan et al.]{K. E. McGowan$^{1}$\thanks{E-mail:
kem@astro.soton.ac.uk}, M. J. Coe$^{1}$, M. P. E. Schurch$^{1}$, 
V. A. McBride$^{1}$, J. L. Galache$^{2}$, 
\newauthor
W. R. T. Edge$^{1}$, R. H. D. Corbet$^{3}$, S. Laycock$^{2}$, 
D. A. H. Buckley$^{4,5}$ \\
$^{1}$School of Physics and Astronomy, Southampton University, Highfield, 
Southampton, SO17 1BJ \\
$^{2}$Harvard-Smithsonian Center for Astrophysics, Cambridge, MA 02138, USA \\
$^{3}$Universities Space Research Association, X-ray Astrophysics Laboratory, 
Mail Code 662, NASA Goddard Space Flight Center, \\
Greenbelt, MD 20771, USA \\
$^{4}$South African Astronomical Observatory, Observatory, 7935, Cape Town, 
South Africa \\
$^{5}$Southern African Large Telescope Foundation, Observatory, 7935, 
Cape Town, South Africa}

\begin{document}

\date{}

\pagerange{\pageref{firstpage}--\pageref{lastpage}} \pubyear{2007}

\maketitle

\label{firstpage}

\begin{abstract}
We have detected 523 sources in a survey of the Small Magellanic Cloud (SMC) 
Wing with {\it Chandra}.  By cross-correlating the X-ray data with optical and 
near-infrared catalogues we have found 300 matches.  Using a technique that 
combines X-ray colours and X-ray to optical flux ratios we have been able to 
assign preliminary classifications to 265 of the objects.  Our identifications 
include four pulsars, one high-mass X-ray binary (HMXB) candidate, 34 stars 
and 185 active galactic nuclei (AGNs).  In addition, we have classified 32 
sources as 'hard' AGNs which are likely absorbed by local gas and dust, and 
nine 'soft' AGNs whose nature is still unclear.  Considering the abundance of 
HMXBs discovered so far in the Bar of the SMC the number that we have detected 
in the Wing is low.
\end{abstract}


\section{Introduction}

Multi-wavelength studies of the Small Magellanic Cloud (SMC) have shown that 
it contains a large number of X-ray binary pulsars.  From analysis of 
H$\alpha$ measurements \citep{ken91} and supernova birth rates \citep{fil98}
the star formation rate (SFR) for the SMC is estimated to lie in the range 
0.04--0.4 $M_{\odot}$ yr$^{-1}$.  \citet{sht05} used these upper and lower SFR 
estimates and the linear relation between the number of high-mass X-ray 
binaries (HMXBs) and the SFR of the host galaxy from \citet{gri03} to predict 
the number of HMXBs expected in the SMC with luminosities 
$\ge 10^{35}$ erg s$^{-1}$.  They found that between 6 and 49 of these systems 
should be present.  Currently $\sim60$ known or probable HMXBs have been 
detected in the SMC \citep[see e.g.][]{hab04,coe05,mcg07}.

It is believed that the considerable number of pulsars can be explained in 
terms of a dramatic phase of star formation, probably related to the most 
recent closest approach of the SMC and the Large Magellanic Cloud 
\citep[LMC;][]{gar96}.  To date most of the X-ray studies of the SMC have 
concentrated on the Bar which has proved to be a significant source of HMXBs.
These systems not only provide an homogeneous sample for study, but also give
direct insights into the history of our neighbouring galaxy as they are tracers
of star formation rates.

Part of the puzzle of the X-ray population of the SMC is the missing or under 
represented components.  In particular, there are no known low-mass X-ray 
binaries (LMXBs) or black hole binaries and only one confirmed supergiant 
X-ray binary detected to date \citep[see also][]{mcb07a}.  A survey of the 
X-ray binary population of the LMC by \citet{neg02} revealed a similar 
distribution (within small number statistics) to that in our galaxy - all 
types were present.  It is therefore important to try and identify the 
``missing'' X-ray binary types in the SMC.

We recently completed the first X-ray survey in the SMC Wing with \chan\ (see 
Section \ref{sect:obs} for more details).  A study of the brightest ($>50$ 
counts) X-ray sources uncovered two new pulsars, and detected two previously 
known pulsars \citep{mcg07}.  In addition to the four pulsars, the sample 
included two foreground stars, 12 probable AGNs and five unclassified sources.
We found that the pulsars had harder spectra than the other bright X-ray 
sources.  In this paper we report on the analysis of the whole survey
and present preliminary classifications for a large fraction of the sources
detected.

\begin{figure}
 \includegraphics[width=84mm]{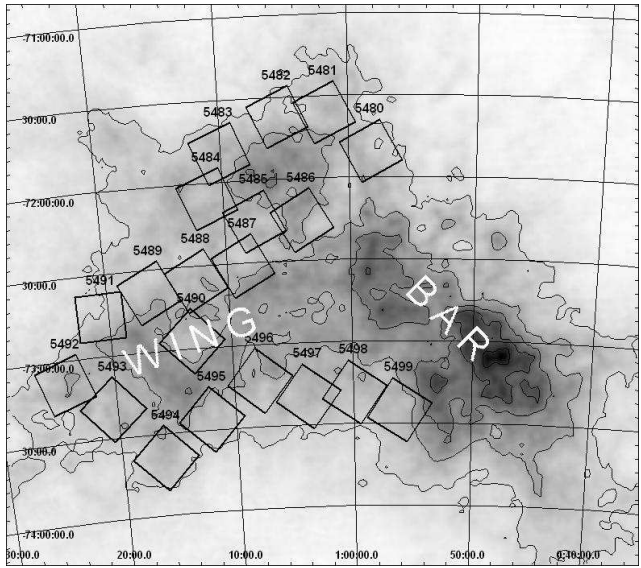}
 \caption{The location of the 20 fields studied by Chandra in this work,
overlaid on a neutral hydrogen density image of the SMC \citep{sta99}.  The
Wing and Bar of the SMC are marked.}
 \label{fig:smc_fields}
\end{figure}

\section{Observations and Data Analysis}
\label{sect:obs}

\citet{coe05} studied the locations of known X-ray pulsars in the SMC Bar and
believed they identified a relationship between the H{\sevensize I} intensity 
distribution and that of the pulsars.  They found that the pulsars seem to lie
in regions of low/medium H{\sevensize I} densities, suggesting that high-mass
star formation is well suited to these densities.  Based on these results 
observations in the Wing of the SMC were made from 2005 July to 2006 March 
with \chan\ (see Figure \ref{fig:smc_fields}).  The survey consisted of 20 
fields, with exposure times ranging from 8.6--10.3 ks.  The observation log is 
presented in Table \ref{tab:log}.  The measurements were performed with the 
standard ACIS-I \citep{gar03} imaging mode configuration which utilises chips 
I0-I3 plus S2 and S3.  

The data were processed using {\sevensize CIAO V}3.3.  We filtered the event 
files to restrict the energy range to 0.5--8.0 keV.
Exposure maps for each field were generated assuming an absorbed power-law 
distribution of source photons with index of 1.6 and neutral hydrogen column 
density of $6 \times 10^{20}$ cm$^{-2}$ \citep{dic90}.  A large number of the 
sources are background active galactic nuclei (AGNs; see Section 
\ref{sect:class}).  The photon index was chosen based on the spectral fitting 
results for the brightest X-ray sources in the survey from \citet{mcg07}.  

\subsection{Source detection}

We searched for sources using the {\sevensize WAVDETECT} tool.  The detection 
algorithm was run on each field using the appropriate exposure map, wavelet 
scales of 1.0, 2.0, 4.0, 8.0 and 16.0 pixels, where a pixel is $0\farcs49$ 
square, and a significance threshold of $10^{-7}$.  The initial wavelet scale 
size is chosen to match the point-spread function (PSF), which has an on-axis 
FWHM of $\sim0\farcs5$.  The size of the PSF is heavily dependent on the 
off-axis angle, increasing to $\sim2\farcs0$ at $\sim6\farcm0$ off-axis.
The choice of significance threshold should yield $\sim1$ false detection over 
a $2048\times2048$ pixel image.  \citet{hor01} have shown that in 
general the counts detected using wavelet analysis agree well with the counts 
determined from aperture photometry.  We converted the observed count rates to 
source fluxes by employing the same absorbed power-law spectrum as above.

\begin{figure}
 \includegraphics[width=84mm]{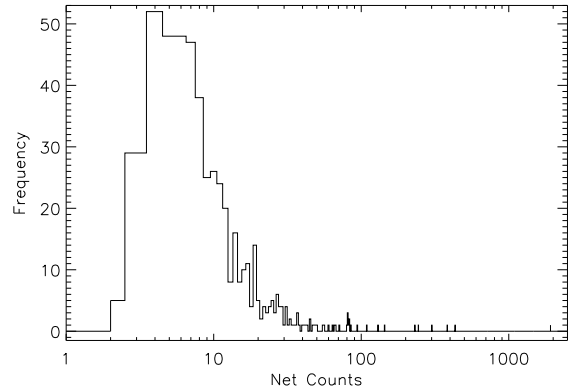}
 \caption{Distribution of net counts for the sources detected in the SMC Wing
survey.}
 \label{fig:counts}
\end{figure}

\section{The SMC Wing Survey Catalogue}
\label{sect:cat}

A total of 523 sources have been detected in the 20 fields.  The number of 
sources detected in each field varies from 16--36 (see Table~\ref{tab:log}), 
with the measured counts per source ranging from 2--1918, with a median of 8 
counts.  The distribution of counts is shown in Figure \ref{fig:counts}.
A sample of the catalogue is presented in Table \ref{tab:cat} (the full 
catalogue is available as Supplementary Material in the electronic edition of 
the journal).  The columns in the table are as follows, the catalogue source 
number, the right ascension and declination taken from the wavelet analysis, 
the error on the source position (see below), the net counts which are the 
total source counts (background subtracted) in the 0.5--8 keV energy band, the 
signal-to-noise of the detection given by the wavelet algorithm, the source 
flux which was determined by converting the observed count rate to a flux 
employing the absorbed power-law spectrum from Section \ref{sect:obs}, the 
median, compressed median and normalised quartile ratio quantile values (see 
Section \ref{sect:quan}), the $V$- and $R$-band magnitudes and the $B-V$ 
colour of the optical counterpart (see Section \ref{sect:opt}), X-ray to 
optical flux ratios based on the $V$- and $R$-band magnitudes, respectively 
(see Section \ref{sect:ratio}), and the preliminary classification for the 
source (see Section \ref{sect:class}).

The significance threshold chosen for our analysis ($10^{-7}$) indicates that 
we should detect $\sim1$ spurious source per field, giving a total of $\sim20$
spurious sources in our catalogue.  We find 17 sources with a signal-to-noise 
of $\leq1.5$ (each has 2--3 net counts) which could be false detections.  For
completeness, we have included these potentially spurious sources in the 
catalogue.

The errors on the source positions were calculated by taking into account the 
properties of the telescope optics and the source brightness 
\citep[see][]{hong05}.  The positional error given in the catalogue is the 
95\% confidence region, combined in quadrature with the boresight error 
($\sim 0.7\arcsec$ at 95\% confidence).  

The median count value, converted to a rate, corresponds roughly to a flux of 
$9\times10^{-15}$ erg cm$^{-2}$ s$^{-1}$.  At a distance to the SMC of 60 kpc
\citep[based on the distance modulus,][]{wes97} this flux corresponds to a
luminosity of $\sim 3.9\times10^{33}$ erg s$^{-1}$.  This limit is adequate to 
detect fainter HMXBs and active LMXBs, but is insufficient for quiescent 
LMXBs which can be as faint as $2\times 10^{30}$ erg s$^{-1}$ \citep{gar01}.

\section{Quantile Analysis}
\label{sect:quan}

We have used the quantile analysis technique of \citet{hong04} to investigate 
the X-ray colours of the sources detected in our survey.  In a traditional 
hardness ratio the photons are split into predefined energy bands.  The 
quantile method divides the photon distribution into a given number of equal 
proportions, where the quantiles are the energy values that mark the boundaries
between consecutive subsets.  This has the advantage, compared to traditional 
hardness ratios, that there is no spectral dependence and a colour can be 
calculated even for sources with very few counts \citep[for more details 
see][]{hong04}.

For each source that has $\geq3$ counts we determine the median and quartiles 
of the photon energy distribution, including background subtraction.  We list 
in Table \ref{tab:cat} the median ($m=Q_{50}$), a compressed median given 
by $\log_{10}(m/1-m)$ and a normalised quartile ratio of $3(Q_{25}/Q_{75})$.
Using the compressed median and the quartile ratio we can construct 
quantile-based colour-colour diagrams (QCCDs).  By generating a quantile-based 
colour-colour diagram \citet{mcg07} were able to investigate the properties of 
the 23 X-ray brightest survey sources.  It was found that the four pulsars 
detected in the SMC Wing Survey lay in a distinct (hard) region on the QCCD. 

\begin{table}
 \begin{minipage}{80mm}
  \caption{Observation Log}
  \label{tab:log}
  \begin{tabular}{@{}cccccc}
  \hline
   Obs  & Date & Central    & Central & Exp  & No. of \\
   ID   &      & RA         & Dec.    &      & Sources \\
        &      & (J2000)    & (J2000) & (ks) &  \\
  \hline
5480 & 2006-02-06 & 00:58:20 & -71:50:27 & 9.59 & 16 \\
5481 & 2006-02-06 & 01:01:54 & -71:35:58 & 9.34 & 31 \\
5482 & 2006-02-06 & 01:05:31 & -71:37:06 & 9.34 & 25 \\
5483 & 2006-02-06 & 01:10:10 & -71:49:29 & 9.34 & 19 \\
5484 & 2006-02-06 & 01:11:20 & -72:05:38 & 9.52 & 30 \\
5485 & 2006-02-08 & 01:07:41 & -72:14:54 & 10.05 & 25 \\
5486 & 2006-02-10 & 01:03:53 & -72:15:06 & 9.83 & 28 \\
5487 & 2006-02-10 & 01:08:47 & -72:30:50 & 9.63 & 29 \\
5488 & 2006-02-12 & 01:12:39 & -72:35:17 & 10.02 & 36 \\
5489 & 2006-02-12 & 01:16:35 & -72:38:16 & 9.63 & 32 \\
5490 & 2006-02-27 & 01:13:21 & -72:57:10 & 10.32 & 25 \\
5491 & 2005-07-24 & 01:20:36 & -72:45:40 & 9.06 & 27 \\
5492 & 2005-08-12 & 01:24:10 & -73:09:02 & 10.06 & 26 \\
5493 & 2006-02-27 & 01:20:28 & -73:19:27 & 9.68 & 23 \\
5494 & 2006-03-01 & 01:16:21 & -73:38:54 & 9.91 & 32 \\
5495 & 2006-03-01 & 01:12:09 & -73:26:10 & 9.63 & 18 \\
5496 & 2006-03-03 & 01:07:55 & -73:13:10 & 9.82 & 24 \\
5497 & 2006-03-03 & 01:03:53 & -73:19:33 & 8.64 & 22 \\
5498 & 2006-03-03 & 00:59:59 & -73:18:34 & 9.63 & 29 \\
5499 & 2006-03-03 & 00:56:10 & -73:25:15 & 9.64 & 26 \\
  \hline								
\end{tabular}							
\end{minipage}							
\end{table}							

\begin{figure}
 \includegraphics[width=84mm]{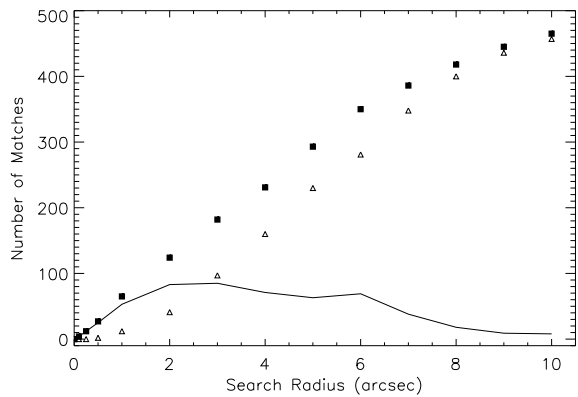}
 \caption{The number of matches as a function of search radius between the
optical/near-infrared catalogues (for a list see Section \ref{sect:opt})
and the SMC Wing X-ray positions (filled squares).  We also show the results 
of correlating the simulated positions (see text for details) with the 
optical/near-infrared catalogues (triangles).  The difference between the
number of real and simulated matches is also plotted (solid line).}
 \label{fig:sim}
\end{figure}

\section{Optical Identifications}
\label{sect:opt}

To search for possible optical counterparts for our \chan\ sources we
cross-correlated the X-ray positions with the following catalogues: Magellanic 
Clouds Photometric Survey: the SMC \citep{zar02}, Guide Star Catalog, 
v.2.2.1 \citep{mor01}, The CCD Survey of the Magellanic Clouds \citep{mas02}, 
USNO-B1.0 Catalog \citep{mon03}, the MACHO online 
database\footnote{http://store.anu.edu.au:3001/cgi-bin/lc.pl} and The 2MASS 
All-Sky Catalog of Point Sources \citep{cut03}.  We chose the search radius 
based on a comparison of a correlation of the real X-ray positions with the 
optical/near-infrared catalogues with a correlation of simulated positions 
with the catalogues, both over a range of search radii 
\citep[see e.g.][]{bar06}.  Only the nearest optical match was taken in each 
case.  The correlations were performed using all of the catalogues given above 
apart from the MACHO catalogue, as it was not possible to automate that 
search.  The simulated positions were generated by mirroring the source 
declinations around an arbitrary declination in the SMC.  This method was 
chosen to try and ensure that the simulated positions lie in similar density 
regions in the SMC as the real positions.

The results of the correlations are shown in Figure \ref{fig:sim}.  The 
difference between the number of real and simulated matches is also plotted.
The value at which the number of real matches is still increasing more rapidly 
than those from the simulated data can be considered as the optimum search 
radius.  Our figure indicates that a radius of $\le 2 \arcsec$ should be used, 
after which the number of real and simulated matches grow at the same rate.  
We have therefore used a search radius of $2 \arcsec$ in our 
cross-correlations, giving us an estimate of the expected number of false 
matches of 33\%.  We note that the choice of search radius is a trade-off
between one that is too small and gives a very conservative number of
matches and one that is too large leading to many spurious matches.  In our
case we find that for four of our sources, all stars, a radius of $2 \arcsec$
is too small and a larger radius of $3 \arcsec$ must be used to obtain a
likely match.

Our calculations of the errors on the source positions (see Section 
\ref{sect:cat}) show that 288 of our sources have uncertainties of
$\leq 2\arcsec$, implying that the chosen search radius is adequate for more 
than half (55\%) of the objects in our survey.  There are 72 sources which have
a positional error of $\leq 1\arcsec$.  For these sources, if the position of 
the nearest optical match differed greatly from the X-ray position we checked
the match by eye to confirm a likely counterpart.  

Out of 523 X-ray sources we find 300 optical and/or near-infrared matches 
within the chosen search radius.  The number of matches shown in Figure 
\ref{fig:sim} is less than this as the matches to the MACHO catalogue are not 
included.  The majority of the fields have a match success of 39--90\%, with 
a median of 67\%.  However, there are three fields 5491, 5492 and 5493 on the 
Eastern edge of the Wing of the SMC (see Figure~\ref{fig:smc_fields}) which 
have very sparse coverage in the optical and near-infrared, resulting in only 
19--26\% of the sources in those fields being matched.

\section{X-ray to Optical Flux Ratios}
\label{sect:ratio}

We have calculated X-ray to optical flux ratios for the sources in our survey
that have optical matches.  It has been shown that such ratios are a good 
discriminator for different classes of objects
\citep[see e.g.][]{mac88,hor01,sht05}.  We have employed two of these kinds of 
ratio, one based on the $V$ magnitude and the other on the $R$ magnitude of the
source.  The former ratio, from \citet{sht05}, is useful for distinguishing 
between HMXBs and stars in the magnitude range $12<V<18$.  In this case stars 
are identified as sources that have $B-V > 0.6$ and $f_{\rm X_{\rm H}}/f_{V} < 
10^{-3}$, where $f_{V} = 8.0 \times 10^{-6} \cdot 10^{-m_{V}/2.5}$ erg 
s$^{-1}$ cm$^{-2}$, and $f_{\rm X_{\rm H}}$ is the flux in the 2--10 keV 
energy band.  The latter ratio can be used to classify AGNs, with typical 
values for these sources lying in the region 
$\log(f_{\rm X_{\rm S}}/f_{R})=0.0\pm 1.0$, where $f_{\rm X_{\rm S}}$ is the 
flux in the 0.5--2 keV range and $\log (f_{\rm X_{\rm S}}/f_{R}) = 
\log f_{\rm X_{\rm S}} + 5.50 + R/2.5$ \citep{hor01}.  Given the measured flux 
for our sources in the 0.5--8 keV energy range we determined the flux in the 
0.5--2 keV and 2--10 keV bands using PIMMS v3.9a, assuming the same absorbed 
power-law spectrum from Section \ref{sect:obs}.

\section{Source Classification}
\label{sect:class}

The X-ray to optical flux ratios, combined with the quantile results, allow 
us to provisionally classify the sources in the SMC Wing Survey.

\subsection{Foreground stars}
\label{sect:stars}

Using the X-ray to optical flux ratio and colour criteria from \citet{sht05} 
we identified a number of stars in the SMC Wing survey.  In addition, a few 
objects with magnitudes brighter than $V=12$ were also classified as stars.
Applying this method we found 16 stars.  These objects are most likely 
foreground stars exhibiting coronal X-ray emission.  We note that the 
$f_{\rm X_{\rm H}}/f_{V}$ values for the sources we have placed in the star 
category range from $1.5 \times 10^{-5} - 0.03$ with a median value of 0.005.  
In the cases where $f_{\rm X_{\rm H}}/f_{V} > 10^{-3}$ the classification was 
made based primarily on the magnitude and colours of the source.  Six of the 
sources we have classified as stars have $f_{\rm X_{\rm H}}/f_{V} \geq 0.01$ 
and fall in the magnitude and colour ranges $V=14.3-17.9$ and $B-V=0.81-1.32$,
respectively.

\subsection{High-mass X-ray binaries}
\label{sect:hmxb}

In \citet{sht05} the authors investigated \xmm\ observations of the SMC, 
mainly located in the Bar, with the purpose of determining HMXB candidates 
in the region surveyed.  A total area of 1.48 deg$^{2}$ was covered with a 
flux limit of $\sim 10^{-14}$ erg s$^{-1}$ cm$^{-2}$ (which corresponds 
to a luminosity of $\sim 4.3 \times 10^{33}$ erg s$^{-1}$ at the distance of 
the SMC).  They cross-correlated the X-ray positions with optical and 
near-infrared catalogues using a search radius of $4\arcsec$.  Any source that 
did not result in an optical match was discarded.  To identify HMXBs they 
required that the magnitude of the optical counterpart lay in the range 
$12.0 < V < 18.0$ and the optical and/or infrared colours were $B-V < 0.6$ and 
$J-K \la 0.1-0.2$, respectively.  They also imposed a limit for the X-ray to 
optical flux ratio, with any source having $f_{\rm X_{\rm H}}/f_{V} < 10^{-3}$ 
being rejected.  Applying these criteria \citet{sht05} found 32 likely HMXBs 
and 18 sources whose nature is uncertain.

Employing the same filters we find four of our sources satisfy the magnitude 
and optical colour criteria (catalogue sources 29, 91, 114 and 193); all of 
which have already been identified as Be X-ray pulsars 
\citep[see][]{mcg07,sch07}.  Out of these four sources, only three have $J-K$ 
values.  The $J-K$ colours for the pulsars are 0.1, 0.3 and 0.6, leading to 
two out of the three sources failing the \citet{sht05} criteria.  However, the 
$J-K$ colours of identified SMC Be X-ray binaries can be shown to lie in a 
much broader range of -0.2 -- $>$0.7 \citep[see e.g.][]{coe05}, which is 
consistent with our sources.  Two other sources meet the magnitude and 
$f_{\rm X_{\rm H}}/f_{V}$ criteria, but they have no colour information 
(catalogue sources 19 and 263).  The $f_{\rm X_{\rm H}}/f_{V}$ ratios for the 
previously classified pulsars lie in the range 0.02--0.45, while the two 
sources without $B-V$ measurements have $f_{\rm X_{\rm H}}/f_{V}=0.01$.  As 
noted above, objects that we have classified as stars can have 
$f_{\rm X_{\rm H}}/f_{V}$ values greater than the limit given in 
\citet{sht05}.  A firm classification for these two sources requires 
additional information (see Section \ref{sect:pattern}).

\subsection{Active galactic nuclei}
\label{sect:agn}

The X-ray to optical flux ratio given in \citet{hor01} was used to establish 
which sources fell in the AGNs category.  This resulted in 51 objects being
classified as AGNs.

\subsection{Low-mass X-ray binaries}
\label{sect:lmxb}

\begin{figure}
 \includegraphics[width=84mm]{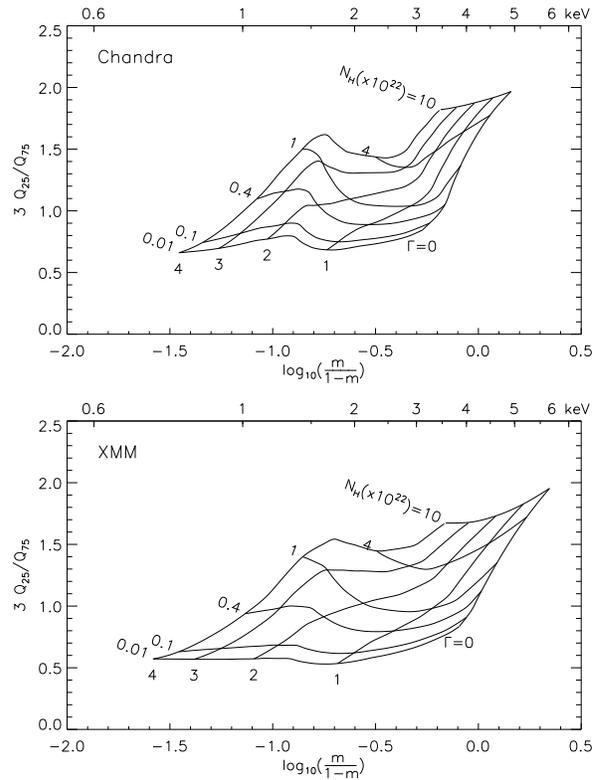}
 \caption{Quantile-based colour-colour diagram based on the median ($m$) and 
the ratio of the two quartiles showing the grid patterns for a power-law model
in a 0.5--8.0 keV range ideal detector for \chan\ (top) and {\it XMM-Newton} 
(bottom).  The top axis shows the median energy values.  The power-law grid
patterns are for $\Gamma=$ 4, 3, 2, 1 and 0 and $N_{\rm H} =$ 10$^{20}$, 
10$^{21}$, $4\times10^{21}$, 10$^{22}$, $4\times10^{22}$ and 10$^{23}$ 
cm$^{-2}$.}
 \label{fig:quan_compare}
\end{figure}

To date no LMXBs have been detected in the SMC and only one, LMC X-2, has been 
detected in the LMC.  In order to investigate the X-ray colours of this source
and compare to the SMC objects, we generated the same quantile values as above 
using an archival {\it XMM-Newton} EPIC-pn observation of LMC X-2 taken on 
2003 April 21.  We find $m=1.63$, $\log_{10}(m/1-m)=-0.75$ and 
$3(Q_{25}/Q_{75})=0.60$ in the 0.5--8.0 keV energy band.  As the quantile 
values are instrument dependent we have generated power-law model grids for a 
range of input spectra using the appropriate response matrix file and 
ancillary response file for each instrument over the energy range of interest
\citep[see][]{hong04}.  Figure \ref{fig:quan_compare} shows how the quantile 
values for \chan\ and {\it XMM-Newton} compare.  We converted the 
{\it XMM-Newton} count rate in the 0.5--8.0 keV energy range to fluxes in the 
0.5--2 and 2--10 keV bands using using PIMMS v3.9a, assuming a power-law index 
of 1.6 and neutral hydrogen column density to the LMC of $6.35 \times 10^{20}$ 
cm$^{-2}$ \citep{dic90}.  Using the $V$ and $R$ magnitudes of LMC X-2 we then 
calculated the two X-ray to optical flux ratios used above, finding 
$f_{\rm X_{\rm H}}/f_{V}=986.54$ and $\log(f_{\rm X_{\rm S}}/f_{R})=3.71$.

\begin{figure}
 \includegraphics[width=84mm]{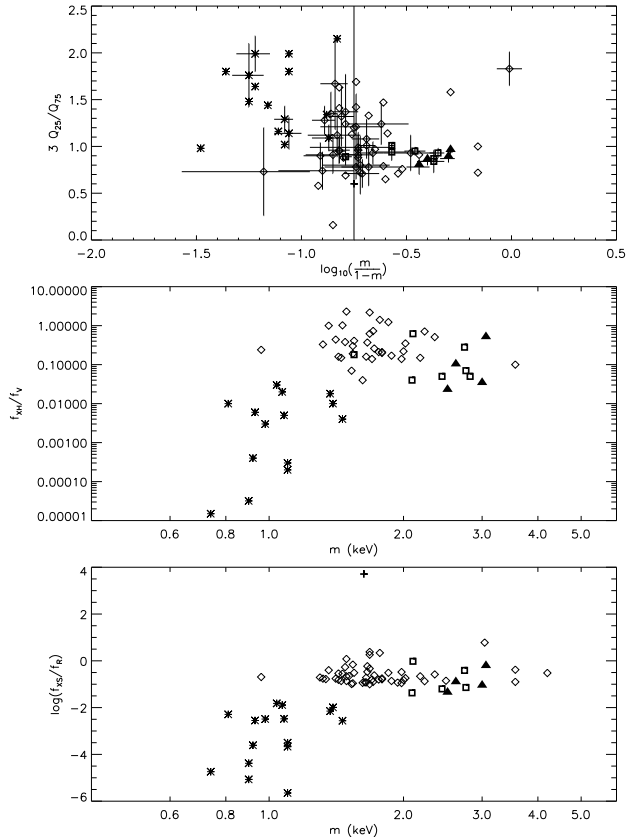}
 \caption{Quantile-based colour-colour diagram (top), quantile median ($m$) 
versus X-ray to $V$ magnitude flux ratio ($f_{\rm X_{\rm H}}/f_{V}$; middle) 
and quantile median ($m$) versus X-ray to $R$ magnitude flux ratio 
($\log(f_{\rm X_{\rm S}}/f_{R})$; bottom) for the sources classified using 
X-ray to optical flux ratios (see Sections \ref{sect:stars}--\ref{sect:lmxb}). 
The objects identified as stars are marked with asterisks, AGNs with diamonds 
and pulsars with filled triangles.  The SMC Bar pulsars from \citet{edg04} are 
also included on the plots, marked with squares, as is the LMXB LMC X-2, 
marked with a cross.  For clarity, we have plotted error bars in the QCCD (top 
panel) only for sources with $>20$ counts.  The position of LMC X-2 falls 
outside of the range of the plot in the middle panel.  These diagrams are used 
as a framework to classify the remaining sources in the survey.}
 \label{fig:pattern}
\end{figure}

\begin{figure}
 \includegraphics[width=84mm]{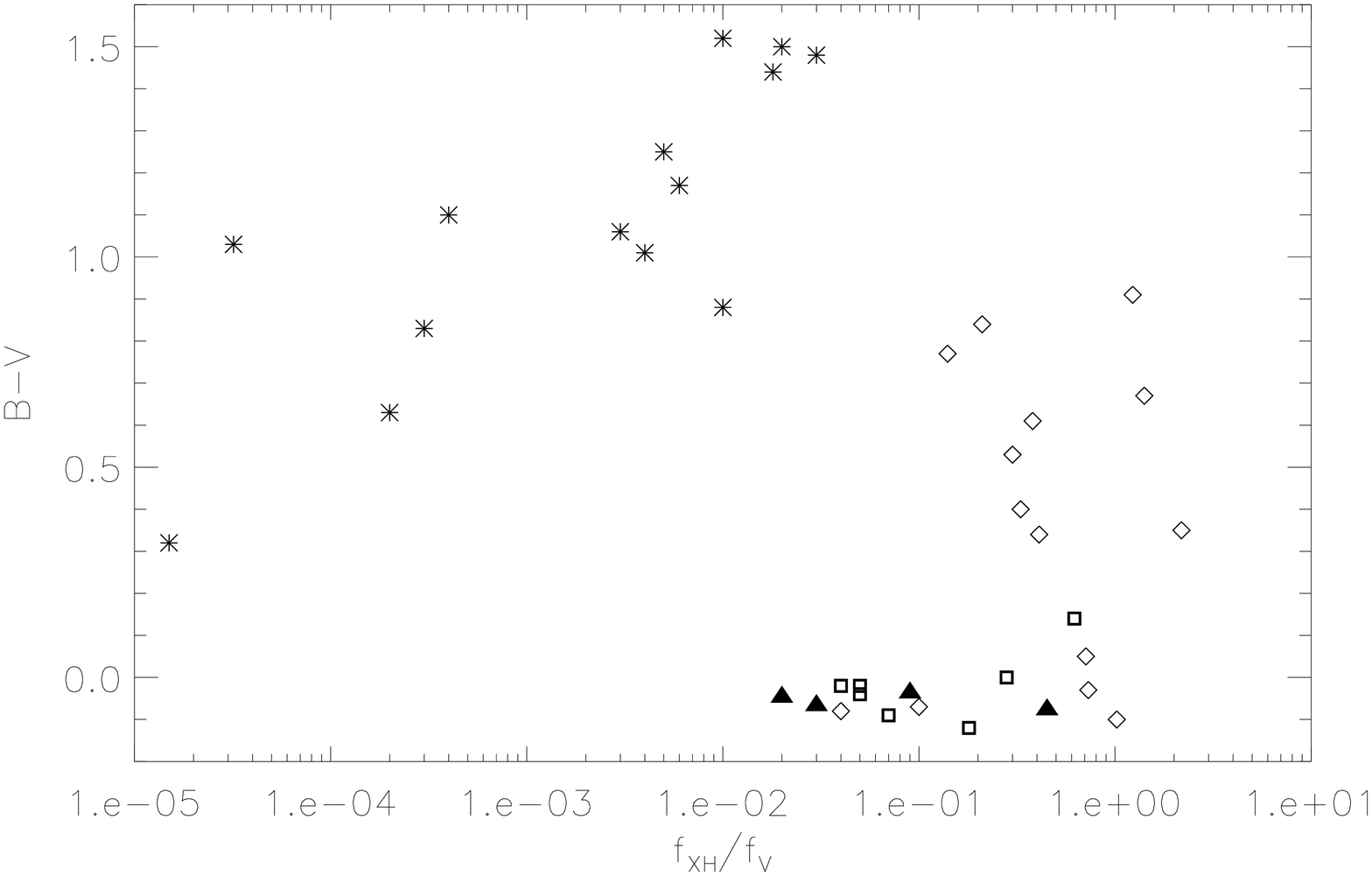}
 \caption{X-ray to $V$ magnitude flux ratio ($f_{\rm X_{\rm H}}/f_{V}$) versus
$B-V$ colour for the sources classified using X-ray to optical flux ratios
(see Sections \ref{sect:stars}--\ref{sect:agn}).  The symbols represent the
same classifications as in Figure \ref{fig:pattern}.}
 \label{fig:colour}
\end{figure}

\subsection{X-ray to optical flux ratios combined with quantile analysis}
\label{sect:pattern}

We created a QCCD for the four confirmed pulsars, 16 stars and 51 AGNs 
classified above (see Figure~\ref{fig:pattern}, top).  We have also included
on the QCCD the SMC Bar pulsars from \citet{edg04} and the LMXB LMC X-2.  As 
can be seen from the figure, while the majority of the stars, AGNs and pulsars 
seem to occupy different parts of the diagram, there is overlap making it 
difficult to classify a source based purely on the QCCD.

For the same sources we also plot the quantile median ($m$) versus X-ray to 
$V$ magnitude flux ratio ($f_{\rm X_{\rm H}}/f_{V}$) and quantile median ($m$) 
versus X-ray to $R$ magnitude flux ratio ($\log(f_{\rm X_{\rm S}}/f_{R})$) in 
Figure~\ref{fig:pattern}, middle and bottom, respectively.  Not all of the 
sources have both $V$ and $R$ magnitudes, with greater coverage in the 
$R$-band.  Presenting the data in this way seems to provide a clearer 
discriminator between the object classes, in particular for the stars and AGNs.
The stars tend to be faint in X-rays and bright in the optical leading to 
small X-ray to optical flux ratios and hence lie in the softer region of the 
diagrams, i.e. they have low quantile median values.  In 
general, the AGNs have quantile median values in the range 1.4--2 keV.  The 
few AGNs that lie in the harder region of the plots are most likely heavily 
absorbed by local dust and gas.  There is also one AGN that falls in the
soft region of the diagrams.  As noted in the previous SMC Wing survey paper 
\citep{mcg07}, the Wing pulsars have relatively hard spectra.

Using these diagrams as a framework we can try and classify the remaining 
sources in the survey.  In particular we can use the above relationships for 
objects that cannot be classified using the appropriate X-ray to optical 
ratios, i.e. AGNs that do not have $R$ magnitudes and stars that do not have 
$V$ magnitudes.  While it would be desirable to be able to classify all of the 
sources based on X-ray data alone, our results show that this is not possible. 
The drawback of our method is that it relies on the optical matches being 
correct.  If there is ambiguity in the object position on the X-ray flux to
optical flux ratio diagrams our classification is made with reference to its
position on the QCCD.

We also show in Figure \ref{fig:colour} the X-ray to $V$ magnitude flux ratio
$f_{\rm X_{\rm H}}/f_{V}$ versus $B-V$ colour for the sources we have 
classified in Sections \ref{sect:stars}--\ref{sect:agn} using the X-ray to 
optical flux ratios combined with the results from the quantile analysis.  
Again, the figure shows a clear division between the stars and the AGNs, and 
the stars and the pulsars.  While this diagram lends emphasis to our 
classification method, many fewer sources have colour information so we find 
that we cannot rely on this as the main tool for our classification.

\begin{figure}
 \includegraphics[width=84mm]{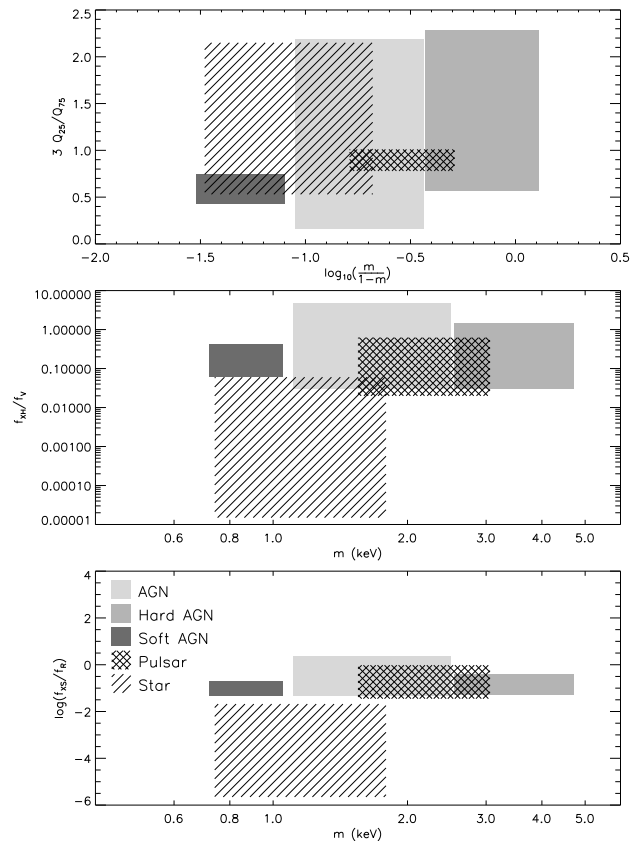}
 \caption{The parameter spaces described by the sources we have classified in 
the SMC Wing survey as stars, AGN, `hard' AGN, `soft' AGN and pulsars 
(see Sections \ref{sect:patt_stars}--\ref{sect:patt_hmxb}).  We have included
the Bar pulsars \citep{edg04} in the pulsar group.  The star category is 
plotted as a hatched region, the AGN in light grey, the `hard' AGN in 
mid-grey, the `soft' AGN in dark grey and the pulsars as a cross-hatched 
region.}
 \label{fig:class_crit}
\end{figure}

\subsubsection{Stars}
\label{sect:patt_stars}

By combining the quantile and X-ray to optical flux ratio data we identify
34 stars in the SMC Wing Survey.  Using the SIMBAD database we have found
the spectral types for four of these sources and find two F, one G and one M 
star.  In two cases we have classified a source as a star based on optical
data alone.  One of the objects flagged as a possible HMXB in Section 
\ref{sect:hmxb} (catalogue source 263) is subsequently identified as a star 
from its position on the combined quantile analysis and flux ratio diagrams.

\subsubsection{AGNs}
\label{sect:patt_agn}

\begin{figure}
 \includegraphics[width=84mm]{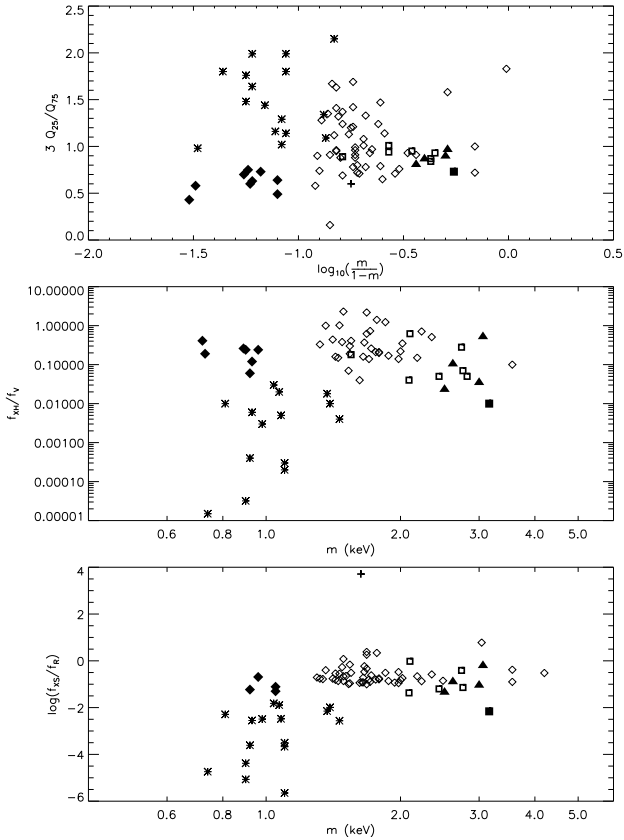}
 \caption{The same diagrams as in Figure \ref{fig:pattern} but with the 
'soft' AGNs and the HMXB candidate included.  The symbols represent the same 
classifications as in Figure \ref{fig:pattern}.  Only the sources classified 
in Sections \ref{sect:stars}--\ref{sect:lmxb} are shown.  The sources 
classified as 'soft' AGNs are marked with filled diamonds, and the HMXB 
candidate is marked with a filled square.}
 \label{fig:patt_sagn_hmxb}
\end{figure}

The majority of the sources in our survey are candidate AGNs.  Based on the 
quantile median ($m$) of an object we have classified the source as either an 
AGN ($1.1<m<2.5$) or a 'hard' AGN ($m>2.5$).  We find 185 of the former and 32 
of the latter.  In addition, we identify a subset of sources which we have 
called 'soft' AGNs which have $m<1.1$ (see Figure~\ref{fig:patt_sagn_hmxb}).  
This sub-class is based on the classification of one of the original sources 
as an AGN (see Section \ref{sect:pattern}) which falls in a region separated 
from the location of the stars and AGNs.  Optical spectroscopy is needed to 
determine the nature of these nine sources, and confirm whether or not they 
are AGNs.

According to the relation from \citet{hor01} the 
$\log(f_{\rm X_{\rm S}}/f_{R})$ ratio for AGNs should be $0\pm 1$.  We find 
that a small fraction of the sources that we have classified as AGNs have 
$\log(f_{\rm X_{\rm S}}/f_{R}) < -1$.  This could indicate 
that we have the wrong optical match, or that the uncertainty on the $R$ 
magnitude is large.  In most of these cases the classification was based on 
the location of the source in the QCCD, supported by the flux ratio plots.

\subsubsection{HMXBs}
\label{sect:patt_hmxb}

Apart from the four already known pulsars we find only one source 
(catalogue source 19, CXOU J005635.0-732631, RA = 00:56:34.96, 
Dec. = -73:26:30.6) that was identified as a HMXB candidate in Section 
\ref{sect:hmxb}.  We note that this object is fainter than the majority of 
HMXBs identified to date, with $V=18.0$ and $R=17.2$.  The other source that 
met the magnitude and $f_{\rm X_{\rm H}}/f_{V}$ criteria (see Section 
\ref{sect:hmxb}) was subsequently identified as a star (see Section 
\ref{sect:patt_stars}).  The combined quantile analysis and flux ratio 
diagrams for the HMXB candidate indicate that it could be categorised as such 
(see Figure \ref{fig:patt_sagn_hmxb}).  However, the presence of a fainter 
object, below the catalogue thresholds, closer to the X-ray position cannot be 
ruled out.  Again, optical spectroscopy, or detection of pulsations, is 
required to determine its true character.

In Figure \ref{fig:class_crit} we plot the parameter spaces described by the
sources we have classified as stars, AGN, `hard' AGN, `soft' AGN and the Bar
and Wing pulsars.

\subsubsection{LMXBs}
\label{sect:patt_lmxb}

We do not find any sources that seem to have the characteristics of a LMXB, but
we cannot rule out the presence of quiescent sources below our detection
threshold of $3.9\times10^{33}$ erg s$^{-1}$.

\subsubsection{Others}

In addition to the classes given above, we find 35 sources that have optical 
matches for which an unambiguous classification is not possible, including  
one source that has only two counts so no quantile information is available. 
It is likely that for a number of these objects the wrong optical match has 
been made leading to a discrepancy in the position of the source on the flux 
ratio plots compared to the QCCD.  We have also not attempted to classify the 
objects that do not have optical counterparts due to the ill defined 
boundaries for different types of sources on the QCCD.

\section{Discussion}
\label{sect:disc}

For the 523 sources detected in the SMC Wing survey we have been able to find 
optical matches for 300 of them, and assign preliminary classifications to
265 objects.  Our classification method has the advantage that it does not 
require optical spectra, however, it still requires optical counterparts to be
identified.  We also note that to classify the remaining 49\% of the survey
deeper optical surveys are needed, and in some cases better coverage of the 
Wing.  

The majority of the Wing sources are found to be AGNs.  In the whole survey we 
only identify four pulsars \citep[see][]{mcg07} and one HMXB candidate, which 
compared to the Bar is a small sample.
The relatively few pulsars detected in the Wing is perhaps not surprising 
given the accepted link between regions of H$\alpha$ and star formation, with 
the main regions of star formation coinciding with the high density H$\alpha$ 
region in the Bar \citep{ken95}.  However, in general, the pulsars we detected 
in the Wing have harder spectra than those in the Bar.  It is also remarkable 
that the only supergiant system so far detected in the SMC, SMC X-1, lies in 
the Wing.  We note that, despite appearances, the SMC is a very three 
dimensional object.  Studies of the Cepheid population by \citet{lan86} have 
revealed that the depth of the SMC is up to 10 times its observed width.  The 
two main structures, the Bar and the Wing, could be separated by 10--20 kpc.  
Could different populations be represented in the two regions?

In the case of the HMXBs if we based our response on the X-ray results alone
we could perhaps draw the conclusion that the sources in the Wing and Bar are 
in fact different.  However, taking into account the optical spectral analysis 
in which the optical counterparts for the pulsars were found to be typical of 
other HMXBs in the SMC \citep{sch07,mcb07b}, different populations seem 
less likely.  This could imply that there is absorption local to the sources
which effects the X-ray spectral results.

There is also the possibility that a greater population of HMXBs does exist in 
the Wing of the SMC, but we were not fortunate enough to catch more than a 
handful of them when they were switched on.  From our studies of 10 years of 
{\it RXTE} data we find that the probability of a Be X-ray transient being in 
an active phase is only, on average, $\sim 10$\% 
\citep[Figure 4.62,][]{gal06}. Quiescent X-ray transients have been detected 
previously in the Milky Way with luminosities $<10^{34}$ erg s$^{-1}$ 
\citep[e.g.][]{neg00,cam02}.  The origin of the quiescent luminosity in Be 
X-ray transients is still under debate, with a number of processes suggested 
to account for the detected emission \citep[see e.g.][]{cam02,kre04}.  The two 
mechanisms detectable from sources located in the SMC are: accretion onto the 
magnetospheric boundary, the propeller regime \citep{ill75,cam00}, and very 
low rate accretion onto the surface of the neutron star, i.e. residual/leaking 
accretion \citep[e.g.][]{ste94}.  The one HMXB candidate that we have 
identified has a luminosity (at the distance to the SMC) of $3.2\times10^{33}$
erg s$^{-1}$ so it could be a quiescent source.

The lack of HMXBs in the Wing indicates that we are looking at an older 
population which is confirmed by optical studies of the star formation history 
of the SMC \citep[e.g.][]{har04}.  In theory this should increase our chances
of detecting LMXBs.  Arguably, LMXBs should be well distributed within the
SMC, i.e. they should lie in the Bar and the Wing, however, deep looks of
the SMC Bar \citep{naz03} have been unsuccessful in detecting any.

The number of LMXBs expected in the SMC is proportional to the total stellar 
mass of the galaxy, resulting in a prediction of only one system with an X-ray 
luminosity of $\geq 10^{35}$ erg s$^{-1}$ \citep[see][]{sht05}.  However, 
\citet{gar01} have shown that quiescent LMXBs can be as faint as 
$2\times 10^{30}$ erg s$^{-1}$.  To go as deep as that is beyond the 
capability of current X-ray telescopes, but in 100 ks it would be possible to 
reach a limit of $\sim 10^{32}$ erg s$^{-1}$, sufficient to detect a sample of 
fainter sources and study their characteristics.  If an observation like this 
were performed in the Wing it could be compared directly with the deep 
exposures of the Bar \citep{naz03,zez05} and help quantify the LMXB population 
in the SMC.

\section*{Acknowledgments}

RHDC and SL acknowledge support from Chandra/NASA grant GO5-6042A/NAS8-03060.
The authors wish to thank JaeSub Hong for making the quantile analysis code 
available. 
This paper utilizes public domain data originally obtained by the MACHO 
Project, whose work was performed under the joint auspices of the U.S. 
Department of Energy, National Nuclear Security Administration by the 
University of California, Lawrence Livermore National Laboratory under 
contract No. W-7405-Eng-48, the National Science Foundation through the Center 
for Particle Astrophysics of the University of California under cooperative 
agreement AST-8809616, and the Mount Stromlo and Siding Spring Observatory, 
part of the Australian National University.
This research has made use of the SIMBAD database, operated at CDS, Strasbourg,
France.  
We thank the referee, John Pye, for useful comments that have 
helped improve the paper.

\begin{table*}
\caption[]{The SMC Wing Survey Catalogue (only the first 20 sources are 
shown).  The full catalogue is available as Supplementary Material in the 
electronic edition of the journal.  The table is ordered in ascending RA.  The 
sources are classified as the following: star, AGN, pulsar, 'AGN h' and 
'AGN s' - hard and soft AGN, respectively (see text for details) and 'HMXB?' - 
HMXB candidate.  The sources with optical matches that were not able to be 
classified are marked with '?'.}\label{tab:cat}
\resizebox{!}{650pt}{ \rotatebox{90}{
  \begin{tabular}{@{}cccccccccccccccc}
\hline
No & RA      &  Dec.   &  Error   &  Net  & S/N &  Flux                                      & $m$ & $\log_{10}(m/1-m)$ & $3Q_{25}/Q_{75}$ & $m_{V}$ & $m_{R}$ & $B-V$ & $f_{\rm X_{\rm H}}/f_{V}$ & $\log(f_{\rm X_{\rm S}}/f_{R})$ & Class \\
   & (J2000) & (J2000) & ($\pm \arcsec$) &  Cts  &     &  ($\times 10^{-14}$ erg cm$^{-2}$ s$^{-1}$) & (keV) &   &     &         &         &       &                 &                 &        \\  
\hline
1 & 00:54:35.47 & -73:19:40.8 & 4.17 &  11 & 3.7   & 2.31 & 1.46 & -0.83 & 1.99 & 19.2 & 18.3	    &   & 0.11 & -1.23 & AGN \\
2 & 00:55:03.67 & -73:21:10.5 & 2.43 &   9 & 4.2   & 1.15 & 1.85 & -0.66 & 1.05 & 18.8 & 	    & 0.18  & 0.04 &  & AGN \\
3 & 00:55:16.97 & -73:23:49.6 & 1.17 &  12 & 5.8   & 1.30 & 1.81 & -0.67 & 0.62 & 19.4 & 19.2 &   & 0.08 & -1.12 & AGN \\
4 & 00:55:17.38 & -73:30:08.5 & 3.13 &   6 & 2.9   & 0.71 & 1.05 & -1.10 & 0.64 &  & 19.4 &   &  & -1.30 & AGN s \\
5 & 00:55:24.34 & -73:31:11.0 & 3.97 &   6 & 2.8   & 0.74 & 0.76 & -1.45 & 0.70 &      &      &     &	&  & \\
6 & 00:55:26.45 & -73:34:37.6 & 11.16 &  6 & 2.5   & 0.97 & 1.52 & -0.80 & 0.64 & 20.3 &  & 0.79  & 0.13 &  & AGN \\
7 & 00:55:32.50 & -73:23:16.7 & 2.51 &   2 & 1.0   & 0.21 &  -   &   -   &  -   & 20.8 & 20.6  &     & 0.04 & -1.35 & ? \\
8 & 00:55:32.94 & -73:31:14.4 & 2.69 &   9 & 4.1   & 1.36 & 2.17 & -0.54 & 0.93 &      &      &     &	&  & \\
9 & 00:55:33.34 & -73:18:20.2 & 2.01 &  19 & 8.1   & 2.51 & 1.79 & -0.68 & 1.33 & 19.8 & 19.4 &   & 0.21 & -0.75 & AGN \\
10 & 00:55:45.25 & -73:24:45.3 & 0.96 &   6 & 3.1   & 0.70 & 1.26 & -0.95 & 1.22 & 18.7 & 18.3 &    & 0.02 & -1.75 & star \\
11 & 00:55:51.54 & -73:31:10.1 & 0.88 & 231 & 74.2  & 27.4 & 1.46 & -0.83 & 1.10 & 18.4 &  & 0.61   & 0.64 &  & AGN \\
12 & 00:56:03.63 & -73:23:24.6 & 0.85 &  15 & 7.7   & 1.55 & 1.54 & -0.79 & 1.24 & 20.2 & 19.4 &    & 0.19 & -0.96 & AGN \\
13 & 00:56:05.22 & -73:29:45.2 & 2.27 &   4 & 2.1   & 0.55 & 3.05 & -0.29 & 1.79 &  &  &   &  &  &  \\
14 & 00:56:11.60 & -73:28:51.7 & 1.45 &   3 & 1.6   & 0.33 & 1.52 & -0.80 & 2.29 &      & 	    &       &	&  & \\
15 & 00:56:12.39 & -73:26:30.4 & 1.49 &   4 & 2.1   & 1.08 & 1.68 & -0.73 & 1.36 &      &      &     &  &  &  \\
16 & 00:56:16.36 & -73:25:19.8 & 0.86 &   7 & 3.7   & 0.73 & 1.36 & -0.89 & 1.89 & 20.4 &  & 0.26   & 0.11 &  & AGN \\
17 & 00:56:20.27 & -73:24:25.9 & 0.99 &   4 & 2.1   & 0.42 & 1.87 & -0.65 & 0.83 & 19.9 & 19.9	    &   & 0.04 & -1.33 & ? \\
18 & 00:56:24.09 & -73:25:06.2 & 0.88 &   7 & 3.6   & 0.73 & 2.22 & -0.53 & 0.69 & 21.0 &  & 0.40   & 0.19 &  & AGN \\
19 & 00:56:34.96 & -73:26:30.6 & 0.96 &   7 & 3.6   & 0.75 & 3.16 & -0.26 & 0.73 & 18.0 & 17.2 &    & 0.01 & -2.16 & HMXB? \\
20 & 00:56:38.64 & -73:28:56.3 & 1.75 &   5 & 2.6   & 0.66 & 0.80 & -1.38 & 0.64 &      &      &     &	&  & \\
\hline
\end{tabular}
}}
\end{table*}

\label{lastpage}

\end{document}